\documentclass[usenatbib]{article}
\usepackage{times,graphicx,natbib,a4wide}

\begin{document}
	
\title{Virulence as a Model for Interplanetary and Interstellar Colonisation - Parasitism or Mutualism?}
\author{Jonathan Starling$^1$ and Duncan H. Forgan$^2$}
\maketitle

\noindent $^1$SAC, Kings Buildings, West Mains Rd, Edinburgh, EH9 6GU, UK  \\
\noindent $^2$Scottish Universities Physics Alliance (SUPA), Institute for Astronomy, University of Edinburgh, Blackford Hill, Edinburgh, EH9 3HJ, UK \\
\noindent \textbf{Word Count: 5,048} \\

\noindent \textbf{Direct Correspondence to:} Jonathan Starling
\textbf{Email:} JMS26@hw.ac.uk

% \newpage

\begin{abstract}

In the light of current scientific assessments of human-induced climate change, we investigate an  experimental model to inform how resource-use strategies may influence interplanetary and interstellar colonisation by intelligent civilisations.  In doing so, we seek to provide an additional aspect for refining the famed Fermi Paradox.  The model described is necessarily simplistic, and the intent is to simply obtain some general insights to inform and inspire additional models.  We model the relationship between an intelligent civilisation and its host planet as symbiotic, where the the relationship between the symbiont and the host species (the civilisation and the planets ecology, respectively) determines the fitness and ultimate survival of both organisms.  

We perform a series of Monte Carlo Realisation simulations, where civilisations pursue a variety of different relationships/strategies with their host planet, from mutualism to parasitism, and can consequently 'infect' other planets/hosts.  We find that parasitic civilisations are generally less effective at survival than mutualist civilisations, provided that interstellar colonisation is inefficient (the maximum velocity of colonisation/infection is low).  However, as the colonisation velocity is increased, the strategy of parasitism becomes more successful, until they dominate the 'population'.  This is in accordance with predictions based on island biogeography and r/K selection theory.  While heavily assumption dependent, we contend that this provides a fertile approach for further application of insights from theoretical ecology for extraterrestrial colonisation - while also potentially offering insights for understanding the human-Earth relationship and the potential for extraterrestrial human colonisation.

\end{abstract}

\section{Introduction}

The Fermi Paradox suggests that the timescales required for the
development of an extraterrestrial civilisation capable of
interstellar colonisation are short compared to either the age of the
Earth or the age of the Galaxy.  Consequently, it should be evident
that the Galaxy is teeming with intelligent life forms.  Their
apparent absence, despite its high probability (by Fermi's reasoning)
led to the question ``where are they?''. 

Numerous hypotheses have been developed to explain the Fermi Paradox.
We will focus on a variant of the so-called Sustainability Solution
\citep{VonHoerner1975, Haqq2009}, which argues that rapid interstellar
colonisation may not be sustainable, i.e. extraterrestrial
civilisations which adopt a rapid colonisation strategy will
eventually fail.  Alternatively, adopting a more sustainable approach
to colonisation leads to a much slower rate of interstellar
colonisation, if at all.

Determining the validity of the Sustainability Solution requires us to
characterise the relationship between an intelligent technological
civilisation and its environment.  While it is impossible to accurately predict the behaviour of other intelligent civilisations when we have no proof of their existence, we can adopt simple models based on known terrestrial ecology to investigate basic behavioural strategies which may be applicable.

In this paper we demonstrate the potential utility of adapting biological and ecological theories to modelling interplanetary and
  interstellar colonisation.  This approach is
  speculative, and not intended as a comprehensive model of such colonisation, but rather a 
  simplified model (limited by various factors and assumptions) with the intention of demonstrating the utility of such an approach for future research.  We feel that there is merit in using models of this type to frame and restrict speculation on our own species' potential extra-terrestrial expansion as well as that of other species.

We construct our model from the following concepts: symbiosis, which
describes inter-species relationships on Earth; \textbf{r/K} selection
theory, which describes species-environment relationships in island
biogeography; and the more controversial Gaian theories which model
colonisation of planets by life as an infectious/reproductive process
occuring between pairs of super-organisms.  The consideration of these
three concepts leads us to a model in which intelligent civilisations
pursue a symbiotic relationship with their host planet.  The nature of
this relationship will inform the colonisation strategy.  Further, the Galactic population of stars and planets will have important
environmental effects, rewarding and punishing certain strategies.

To this end, we have performed Monte Carlo Realisation simulations of
the Milky Way, where civilisations grow and evolve from
non-intelligent organisms in and amongst a population of stars and
planets that are statistically representative.  These civilisations
are assigned a variety of colonisation strategies, and subsequently display varying degrees of success in interplanetary and
interstellar colonisation.  This will allow us to assess what
conditions the civilisations will need to satisfy  to be successful in
the Milky Way (given the
assumptions made in these simplified models).  More specifically, we
are interested in answering two key questions: \emph{What colonisation
  strategies are more successful in the Milky Way? How is this result
  altered when the efficiency of interstellar colonisation changes?}

While the approach taken in this paper is necessarily dependent on a
number of simplifications and assumptions, it does provide some useful
insight into how extra-terrestrial colonization may be informed from
theoretical ecology.  Civilisations are no doubt much more complicated
than was replicable in this exercise; our intention here has not been
to illustrate a mechanistic approach to colonization, but to
investigate what constraints may face colonization as well as
indicating the choices that civilizations may have to make in relation
to their ``host'' planet.  Also, note that we do not consider the consequences of unmanned exploration of the Galaxy, e.g. through the use of fleets of interstellar probes \citep{Bjork2007,Cotta2009,Cartin2013} or swarms of self-replicating probes \citep{Freitas1983, wiley2011,Nicholson2013}.  While the growth of a population of entities via bifurcation has clear biological analogues, this is somewhat outside the scope of this investigation.

The paper is organised thus: in section \ref{sec:virulence} we discuss
the theories of Gaian reproduction, \textbf{r/K} selection and
virulence; in section \ref{sec:code} we discuss the numerical
apparatus employed in this work; in section \ref{sec:results} we
present the results of our simulations, and in sections
\ref{sec:discussion} \& \ref{sec:conclusions} we discuss the
implications of these numerical results and draw our conclusions. 

\section{Gaian Reproduction, r/K selection and Virulence \label{sec:virulence}}

According to the Gaia hypothesis \citep{Lovelock2000} the Earth, with
its biosphere, geosphere and atmosphere, is seen as a complex
interacting system, similar in behaviour to a super-organism.  The
idea of Gaian Reproduction is a subset of this hypothesis, and argues
that the Earth (or any similarly life-sustaining planet) could ``reproduce''
through the transference of life from one planet to another.  This
could be achieved either through active space colonisation by
intelligent and technologically capable life, or through passive
colonisation, where microbial life is accidentally transported to
another nearby planet through a meteorite impact and resulting debris,
usually referred to as panspermia (e.g. \citealt{Wallis2004}).  Active
colonisation could take the form of either terraforming nearby
suitable planetoids, or the discovery and settling of already
habitable planets (or both).   

Gaian reproduction theory espouses two reproduction strategies: the
\emph{sprint} and the \emph{marathon}.  Under a sprint approach,
intelligent life pursues an unsustainable exploitation of the host
planet's resources, using these resources to quickly develop the means
to colonise new planets.  This unsustainable exploitation could lead
either to a collapse of the relevant civilisation, or the collapse of
the biosphere itself.  Under the marathon approach, intelligent life
develops an ecologically sustainable civilisation.  The result is a
longer ``gestation'' period before the civilisation develops the
capacity to colonise new hosts.  It also allows for multiple waves of
colonisation and reduces the risk of host death (an eventuality now
limited to either extra-solar events or the death of the host star).

The two reproductive approaches of sprint and marathon are related to
that of \textbf{r/K} selection theory, developed by the field of
island biogeography as a model of island colonisations by species.
\citet{MacArthur1967} studied the factors relating to the colonisation
of islands by species.  In particular, they focused on the effects of
the island's distance from the species source, and the rate of
extinction for species on arriving to the island, which is itself a
function of area, resources and carrying capacity.  In the process of
this theoretical work they developed what is known today as
\textbf{r/K} selection theory, which describes different life
strategies of species, concerning a trade-off between quantity and
quality of offspring.

Essentially, in \textbf{r}-strategy species the focus is on
reproducing quickly and in the form of many offspring, but with a
reduced metabolic investment (either in nutritional support or
parental care), while \textbf{K}-strategy species have reduced levels
of reproduction and offspring, but a greater metabolic investment,
leading to a greater survival rate of individual offspring.
\textbf{K}-strategy species out-compete \textbf{r}-strategy species
under stable and predictable environments with limited resources and
vice versa.

As such, the sprint strategy of Gaian reproduction is analogous with
\textbf{r}-strategies, and \textbf{K}-strategies with the marathon
approach.  It is important to note however that most species exhibit a
continuum between \textbf{r} and \textbf{K}-strategies, with some more
adaptable species, such as humans, being historically able to adopt
both strategies depending on the environmental situation.

We can make a third analogy with concepts related to \emph{virulence}
and \emph{symbiosis}.  Ultimately, symbiosis refers to a close and
long-term interaction of different species, and can have, generally,
three manifestations, that of parasitism, commensalism and mutualism
(\citealt{Boucher1988}; see also \citealt{Odum2005}, who specify eight
in total, including neutralism, competition, protocooperation,
amensalism and predation).  The difference between these three
manifestations is based on the benefit or detriment of the symbiotic
relationship to the fitness of the species involved
\citep{Boucher1988}.  These three forms of symbiosis are described
below:

\begin{itemize}
\item Parasitism: one species (the parasite) benefits while the other
  (the host) loses in fitness.  This relationship is asymmetrical,
  with a benefit to the parasite involving a detriment to the host.
  An example of such a parasitic symbiosis would be fleas, who feed on
  the blood of its host, reducing its fitness. 
\item Commensalism: one species (the commensal) benefits while the
  other (the host) neither benefits nor is negatively affected by the
  relationship. 
\item Mutualism: both species, the symbiont and the host, benefit from
  the relationship.  An example of a mutualism is that of corals and
  zooanthellae; the coral serves as a host to the zooanthellae,
  providing the symbiont with nutrients and shelter, while the
  symbiont provides the host with photosynthetic energy \citep{Smith1987}. 
\end{itemize}

Virulence relates to the both the degree of damage (or loss of
fitness) incurred to the host by the parasite in question, as well as
the rate of infection from one host to another.  The greater the
damage a parasite causes its host, the greater the risk of the host
dying as a result, and in the process, the lesser the chance of the
parasite of infecting a new host.  Only if there are multiple hosts,
in close proximity, and the ability to transmit from one host to
another is high, does a parasite benefit from an \textbf{r} strategy
of virulence.  If the number of potential hosts is small, or the
ability to transmit to another host is low, then the optimal level of
virulence for a parasite is to adopt a \textbf{K} strategy, leading to
reduced damage to the host.

There exist a number of mathematical models for virulence, of which the below may be considered a general standard \citep{Ebert1996,Weiss2002}:

\begin{equation} \Phi = \frac{\beta(N)}{\mu + \alpha + \nu} \end{equation}

\noindent Where:

\noindent $\Phi$ Symbiont fitness \\
\noindent $\beta$ Rate at which an infected host transmits the parasite \\
\noindent $N$ Host density \\
\noindent $\mu$ Death rate of uninfected host \\
\noindent $\alpha$ The parasite induced mortality rate \\
\noindent $\nu$ Host recovery rate \\

\noindent  We will adopt this model of virulence, where intelligent
civilisations are the ``symbionts'' of their host planets.  The
civilisations' behaviour is described as a continuum between mutualist
and parasitic strategies (or equivalently between \\
{K} and
\textbf{r} strategies).  We should therefore expect that if there are
many available nearby planets as hosts, \textbf{r} strategies will be
more successful, whereas \textbf{K} strategies will dominate if host
planets are not in great supply.  Exactly how these strategies are
implemented in our numerical simulations is discussed in more detail
below. 

\section{Numerical Methods \label{sec:code}}

\noindent To model the growth and evolution of intelligent
civilisations in the Galaxy, we use the Monte Carlo Realisation
techniques as described in \citet{mcseti1,mcseti2}.  In summary, the
method generates a synthetic Galaxy of $N_{stars}$ stars, each with
their own stellar properties randomly sampled from statistical
distributions, such as mass \citep{IMF}, age \citep{Rocha_Pinto_SFH},
chemical composition \citep{Rocha_Pinto_AMR}, location in the Galaxy
\citep{Ostlie_and_Caroll}, etc. Planetary systems are then generated
around some of these stars (depending on their chemical composition,
see \citealt{Wyatt_z}), and life is allowed to evolve in these planets
according to some hypothesis of origin - for example, planets that
exist within the continuous habitable zone of their parent star will
be inhabited (more details of the continuous habitable zone can be
found in \citealt{mcseti2}).  This life is allowed to evolve using
stochastic equations, which account for the possibility of ``resetting
events'' (such as asteroid impacts or local supernovae), which may
impede or completely destroy life on any planet (cf
\citealt{Annis,Vukotic_and_Cirkovic_07}).  Life which survives to
become intelligent undergoes a second phase of danger in which the
civilisation itself may be the architect of its demise.  Civilisations
which do not destroy themselves go on to colonise other planets (and
in the case of this work, planets around other stars).   

The end result is a mock Galaxy with billions of stars and planets,
containing a population of intelligent civilisations, which is to some
degree statistically representative of the Milky Way. To quantify
random sampling errors, this process is repeated many times: this
allows an estimation of the sample mean and sample standard deviation
of the output variables obtained.  Details of this sampling method can
be found in \citep{mcseti1}. 

\subsection{Modelling Civilisation ``Virulence''}

As we are now attempting to model the symbiotic relationship between
intelligent civilisations and their host planets, we must make some
modifications to the method.  Firstly, each civilisation is assigned a
virulence parameter, $\tilde{\alpha}$, which is a normalised version
of the $\alpha$ parameter discussed in the previous section.
$\tilde{\alpha}$ ranges from -1 to +1, -1 being a highly mutualist
civilisation and +1 being highly parasitic, with 0 representing
commensalist civilisations. 

Inhabited planets each possess their own intelligence timescale
$t_{int}$ (i.e. the time it takes for non-intelligent life to become
an intelligent technological civilisation), a maximum habitability
timescale $t_{max}$ (defined by taking the minimum of two timescales -
the star's lifetime $t_{ms}$, and the timescale on which the planet
moves out of the stellar habitable zone $t_{HZ}$), and a total number
of resetting events the planet suffers, $N_{resets}$.  We can use
these variables to construct a normalised host recovery rate
$\tilde{\nu}$:

\begin{equation} \tilde{\nu} = \left(\frac{t_{max}}{t_{int}}\right) N_{resets} \end{equation}

\noindent As we model the extinguishing of life on planets without
intelligence (see \citealt{mcseti2} for details) we implicitly
incorporate the ``uninfected mortality rate'' into our calculations,
hence we do not construct a corresponding $\tilde{\mu}$.

We assume that parasitic civilisations are more likely to be
self-destructing in their early ``fledgling'' stages (which occur
before civilisations are sufficiently advanced to begin colonisation).
To this end, we assign a probability of self-destruction to each
civilisation: 

\begin{equation} P_{destroy} = MIN \left(\frac{1}{\tilde{\nu}(1-\tilde{\alpha})}, 1.0\right). \end{equation}

\subsection{Interplanetary Colonisation}

\noindent If civilisations succeed in becoming advanced, they can attempt to colonise the planets in their system.  The reproductive parameter $\tilde{\beta}$ depends on the properties of the host planet and the planet to be colonised

\begin{equation} \tilde{\beta} = \frac{(1.0+\tilde{\alpha})}{\Delta r\sqrt{M_{p,home}M_{p,col}}}. \end{equation}

\noindent $\Delta r$ is the distance between the two planets and
$[M_{p,home},M_{p,col}]$ are the masses of the civilisation's home
planet and destination planet respectively.  We are interested in the
escape velocity of both planets, hence the square root dependence on
planetary mass.  This function is normalised by the maximum possible
separation and planet mass allowed in the simulation (40 AU,
corresponding to Pluto's orbit, and 10 Jupiter masses respectively). 

The probability of reproduction, for a given planet-pair, is then

\begin{equation} \tilde{\Phi} = \frac{\tilde{\beta}}{\tilde{\nu} + \tilde{\alpha}} \end{equation}

\noindent Using this, the code stochastically reproduces
\emph{interplanetary} colonisation behaviour for every civilisation.
Reproduction will occur where the conditions are most favourable -
i.e. the civilisations are more virulent, the host's recovery rate is
rapid, and the planets are more easily accessible according to their
escape velocity. 

\subsection{Interstellar Colonisation}

We also model a \emph{limited} form of interstellar colonisation,
using a form of revision.  Consider the total civilisation population
over all time for one realisation, without interstellar colonisation.
If we allow the parasitical civilisations (that is, those with
$\tilde{\alpha}>0$) to send out colonising parties into the Galaxy at
some maximum fraction of lightspeed $\chi$, where more virulent
civilisations will travel at greater speed:  

\begin{equation} v_{col} = \tilde{\alpha} \chi c, \end{equation}

\noindent then we can calculate when parasites will arrive at planets
inhabited by intelligent life.  If they arrive before life has become
intelligent, the planet can be colonised by the parasites.  We assume
for simplicity that if parasites arrive after intelligent life evolves
on a planet, then the parasites will not attempt to colonise it.  We
also do not model the colonisation of completely uninhabited worlds,
or the subsequent secondary colonisation that occurs when a colony
decides to begin its own colonisation missions (although this is
obviously of interest for future work). 

By performing this revision in chronological order, we can see which
worlds will eventually contain parasitical civilisations (even if they
did not originally host them).  We can also investigate what the
minimum value of $\chi$ must be for parasitism to be the most
favourable strategy for civilisations to adopt. 

\section{Results \label{sec:results}}

\noindent To constrain the models correctly, we ran two separate
tests, as the modelling process is better equipped to deal with
relative differences than absolute values \citep{mcseti2}.  The first
(and main) simulation allows the value of $\tilde{\alpha}$ to vary
uniformly amongst all civilisations in the simulations.  The second is
a control simulation, where the stellar, planetary and biological
parameters remain identical, but $\tilde{\alpha}=0$ and
$P_{destroy}=0.5$  for all civilisations (i.e. we impose a neutral
colonisation strategy and we are ignorant of what causes
self-destruction).  This allows us to confirm what data depends on
civilisation behaviour, and what data depends on the stellar and
planetary parameters of the mock Galaxy. 

\subsection{Interplanetary Colonisation Only}

\begin{figure}
\begin{center}$
\begin{array}{c}
\includegraphics[scale=0.5]{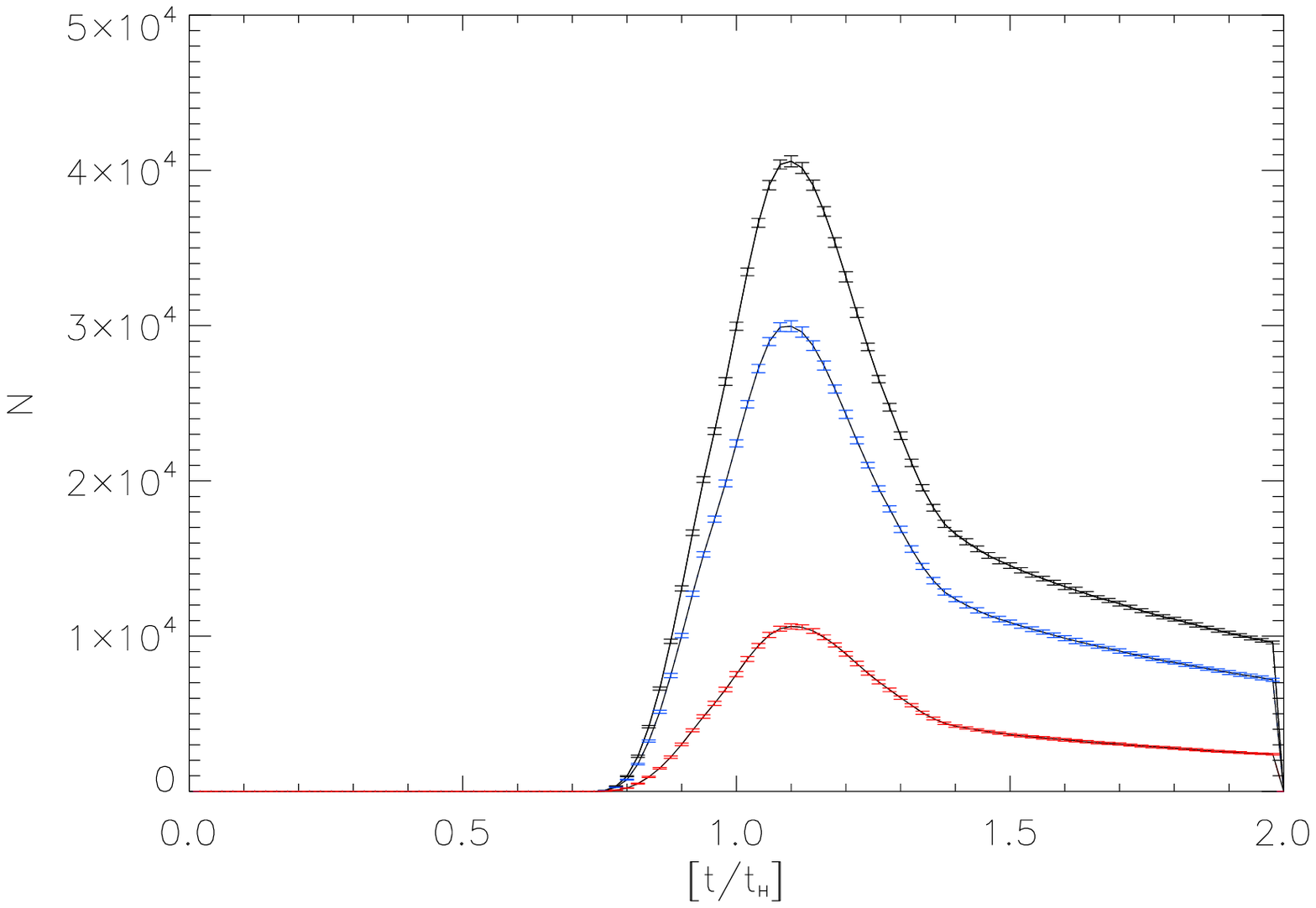} \\
\includegraphics[scale=0.5]{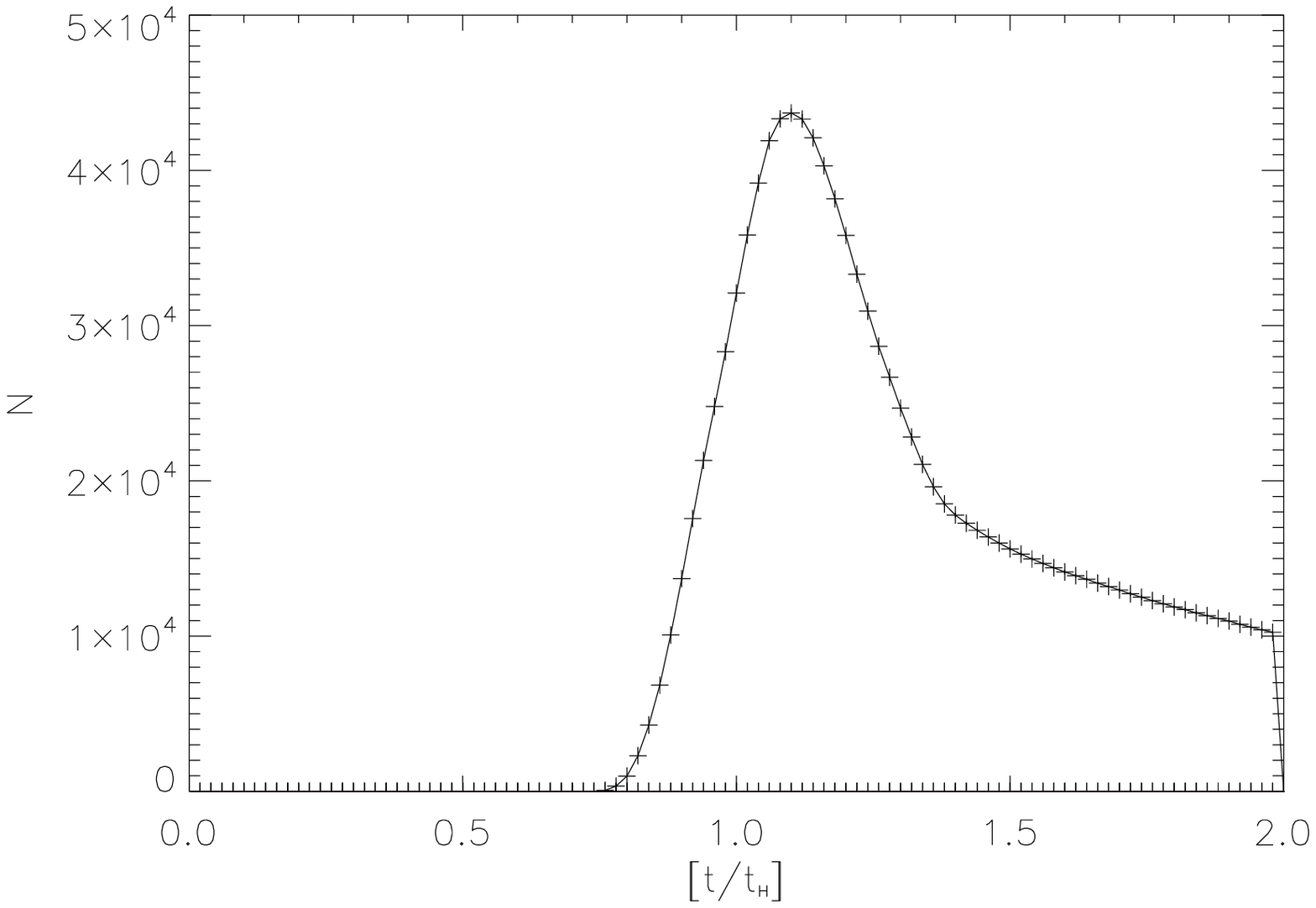} \\
\end{array}$
\caption{Comparing different civilisation strategies.  The top graph
  shows the number of civilisations of each type as a function of
  time, compared to the control simulation (bottom) where
  $\tilde{\alpha}=0$ for all civilisations. The parasitic
  civilisations are displayed in red, the mutualist in blue, and the
  black curve represents all civilisations. The time axis is displayed
  in units of the Hubble Time $t_H$, which is equal to the current age
  of the Universe.   \label{fig:signal}}
\end{center}
\end{figure}

\noindent Figure \ref{fig:signal} displays the number of intelligent
(communicating) civilisations as a function of time for the main
simulation and the control simulation.  As we adopt the Biological
Copernican Principle, and use Earth's biological history as a mean for
other biological histories (see \citealt{mcseti1}) we
self-consistently produce a ``phase transition'' model, where the
number of civilisations increases rapidly after sufficient cosmic time
has elapsed.  We see that in the absence of interstellar colonisation,
mutualism is the more successful strategy.  The peak value of $N$
occurs at $t = 1.1 t_H$ for all civilisation types, and the trend of
total civilisation number is well represented in the control
simulation also (although the peak $N$ is slightly higher, due to the
changed $P_{destroy}$).  All curves show a long tail due to
interplanetary colonisation extending the lifetime of the species to
its maximum value (i.e. the appropriate main sequence lifetime).
These results indicate that limiting the host population to the
planets in one solar system is too restrictive for parasitic species
to operate as successfully as mutualists. 

\begin{figure}
\begin{center}
\includegraphics[scale=0.5]{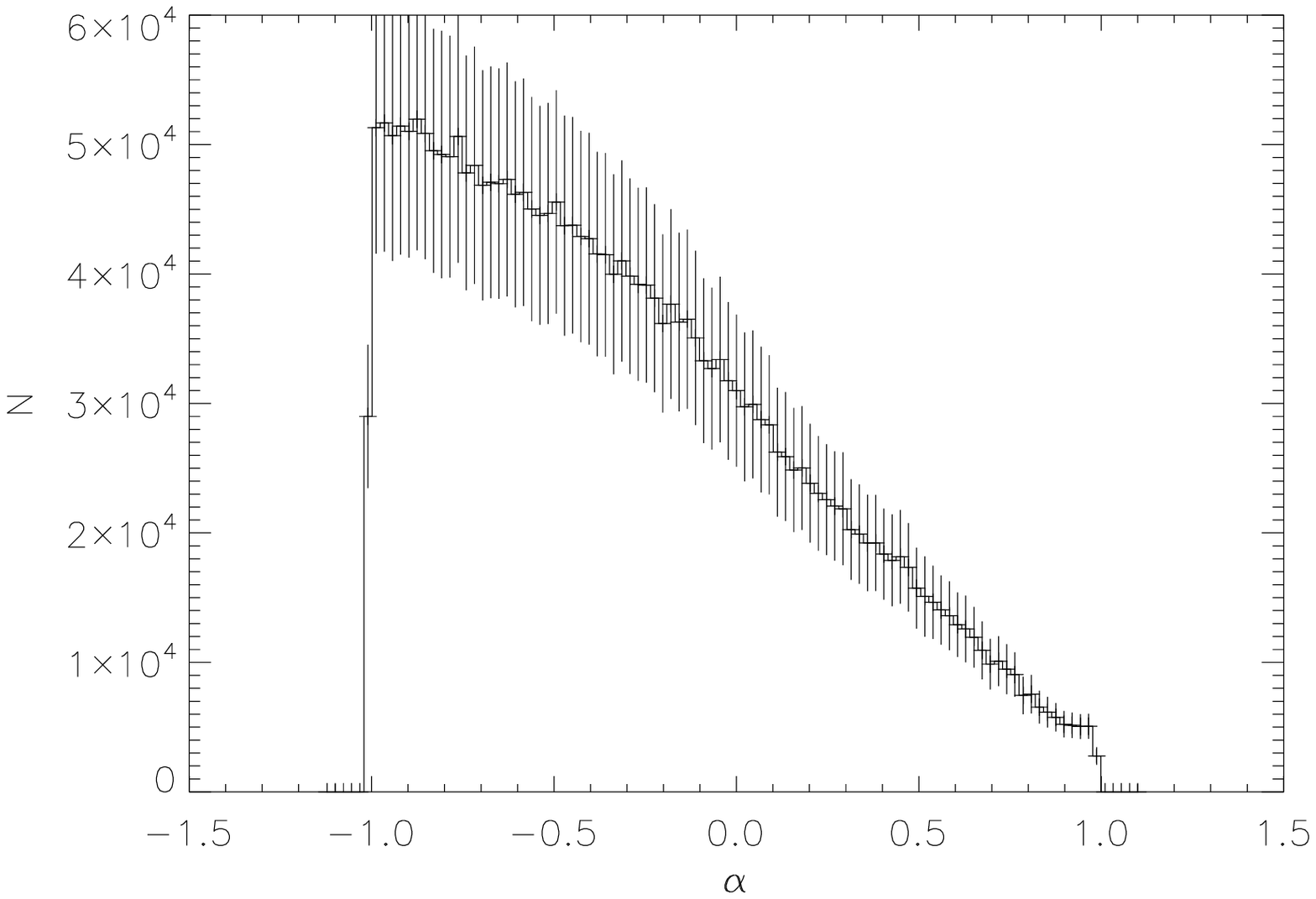} \\
\caption{The distribution of civilisation strategy in the Galaxy
  (without interstellar colonisation).  While initially there is an
  uniform distribution of $\tilde{\alpha}$ amongst all civilisations,
  parasitic civilisations ($\tilde{\alpha}>0$) are punished due to the
  dearth of available host planets - mutualistic civilisations
  ($\tilde{\alpha}<0$) are preferred.  The error bars indicate the
  sample standard deviation taken from 30 distinct realisations.   \label{fig:alpha_nocol}}
\end{center}
\end{figure}

We can see this in Figure \ref{fig:alpha_nocol}.  While the initial
distribution of $\tilde{\alpha}$ is even by construction, the
parasitic civilisations are eventually punished for overexploiting
their host's resources, and hence their total numbers are reduced due
to self-destruction.  Mutualists by comparison fare much better, with
around five times as many extreme mutualists at $\tilde{\alpha}=-1$
exist compared to extreme parasites at $\tilde{\alpha}=1$. 

These results do not account for interstellar colonisation, which
would increase the population of available hosts.  Will this increase
tip the scales in favour of a parasitical strategy? 

\subsection{Interplanetary and Interstellar Colonisation}

\noindent We are free to modify the maximum velocity of colonisation
$v_{max} = \chi c$ (where $c$ is the speed of light \emph{in vacuo}).
For comparison, the Apollo 10 module holds the record for the fastest
manned human vehicle, at 11,082 metres per second, corresponding to
$\chi=3.7 \times 10^{-5}$.  As for unmanned probes, Voyager I is
currently travelling at a speed of approximately 17,062 metres per
second ($\chi=5.7 \times
10^{-5}$)\footnote{http://voyager.jpl.nasa.gov/mission/weekly-reports/}.
It would be reasonable to assume that these are near the lower limit
for interstellar speeds, as the length of the journey would allow for
continued acceleration.  Assuming that interstellar colonists are
limited by the currently known laws of Nature, then $\chi=0.1$ or
higher becomes extremely difficult, and most likely unfeasible.  The
energy required to accelerate one ton of matter to this velocity is
\emph{at least} $4.5 \times 10^{17}$ J, around ten times the current
global consumption of energy\footnote{Statistical Review of World
  Energy 2009, BP. July 31, 2006}.  Assuming that colonisation at this
speed would require large ships with a self-contained population that
can survive for many generations of individuals, the actual energy
budget of colonisation will be several orders of magnitude higher,
limiting the maximum feasible velocity greatly.   

As far as rocket-driven ships are concerned, a maximum velocity of
$\chi=10^{-4}$ is extremely difficult.  The rocket equation has the
following solution for the craft's fraction of mass which is fuel,
$\gamma$: 

\begin{equation} \gamma = 1 - exp\left(\frac{-\chi}{\chi_e}\right) \end{equation}

\noindent Where $\chi_e$ is the effective exhaust velocity.  To
achieve speeds of $\chi=10^{-4}$ with an effective exhaust velocity
equal to that of the Saturn V rocket, the craft's available payload is
only $0.004\%$ of its total mass - i.e. the craft's mass will be
almost entirely composed of fuel.  This would suggest that in the
absence of a propulsion system that does not require propellant to be
stored aboard the craft, we should not expect $\chi$ to be much larger
than $\chi=10^{-4}$.  This assumption is weakened by recent
developments in solar sail technology, (e.g. IKAROS and NanoSail-D) which would not require propellant
storage and would be able to use radiative pressure from nearby stars
to accelerate and decelerate.  However, it is unclear what the maximum
feasible velocity of such an interstellar craft is.  Also, judicious use of slingshot maneouvres \citep{probe_staticbox} could help boost even quite slow craft to large speeds, but only if a large number of maneouvres are available.

\begin{figure}
\begin{center}$
\begin{array}{c}
\includegraphics[scale=0.5]{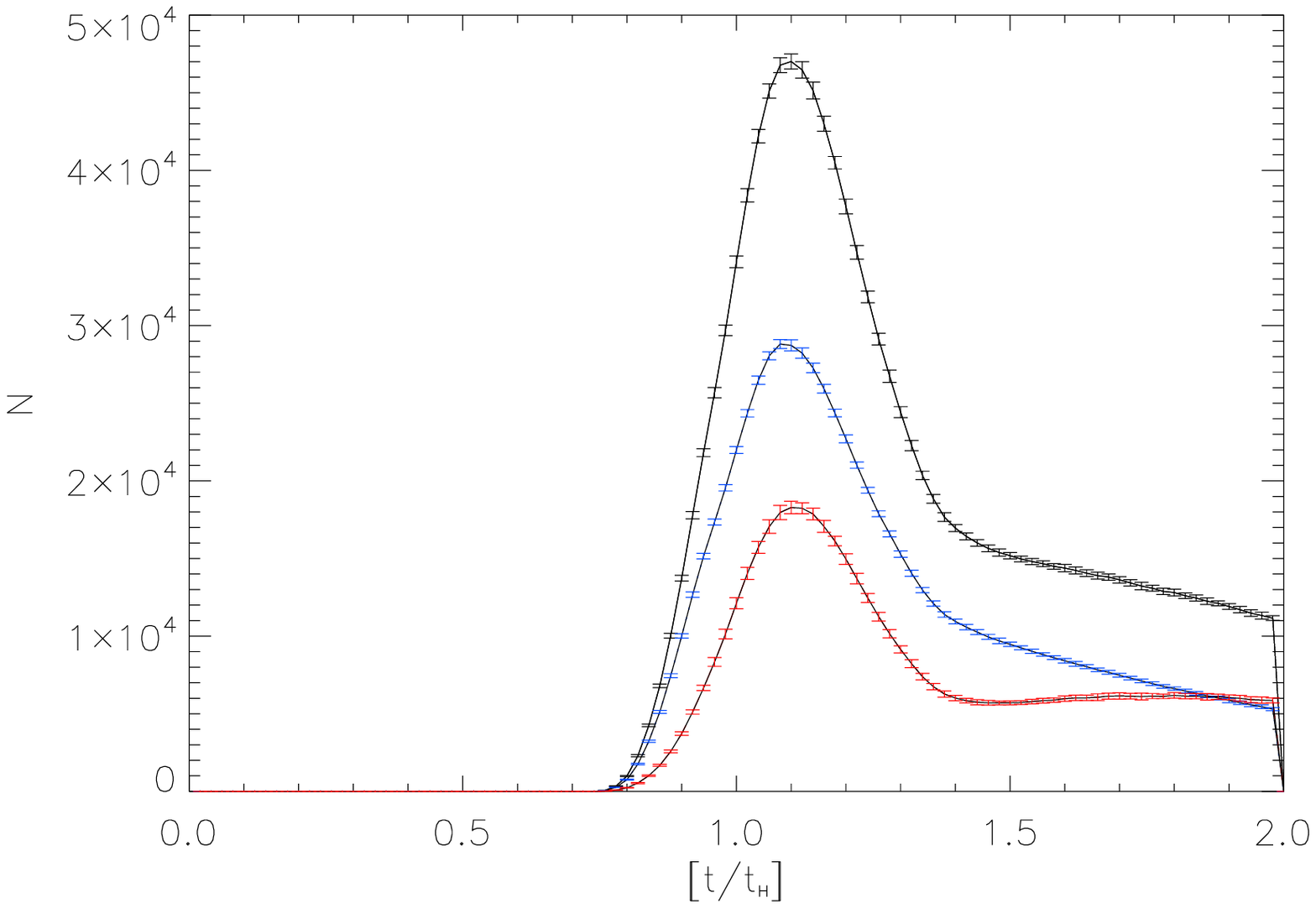} \\
\includegraphics[scale=0.5]{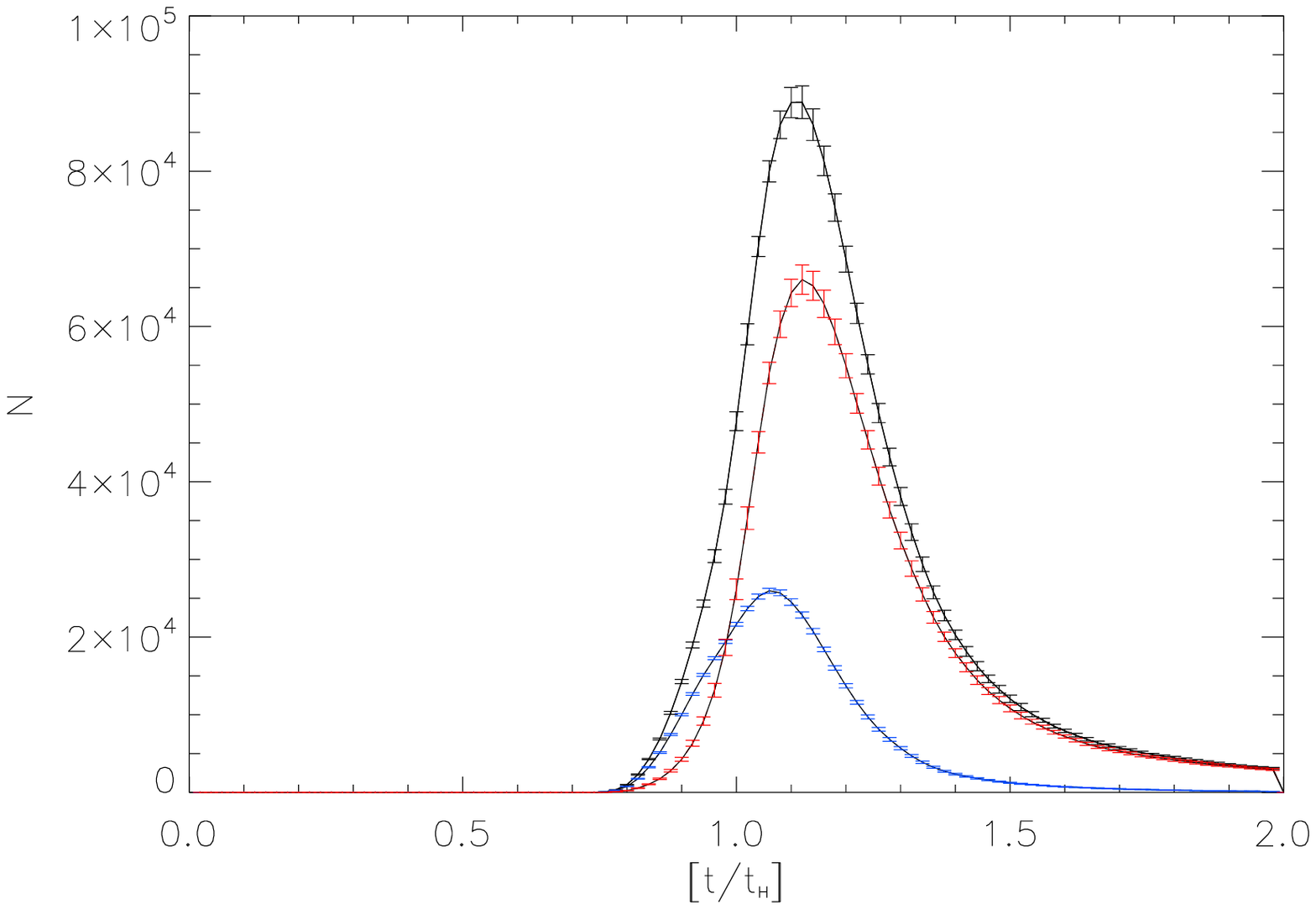} \\
\includegraphics[scale=0.5]{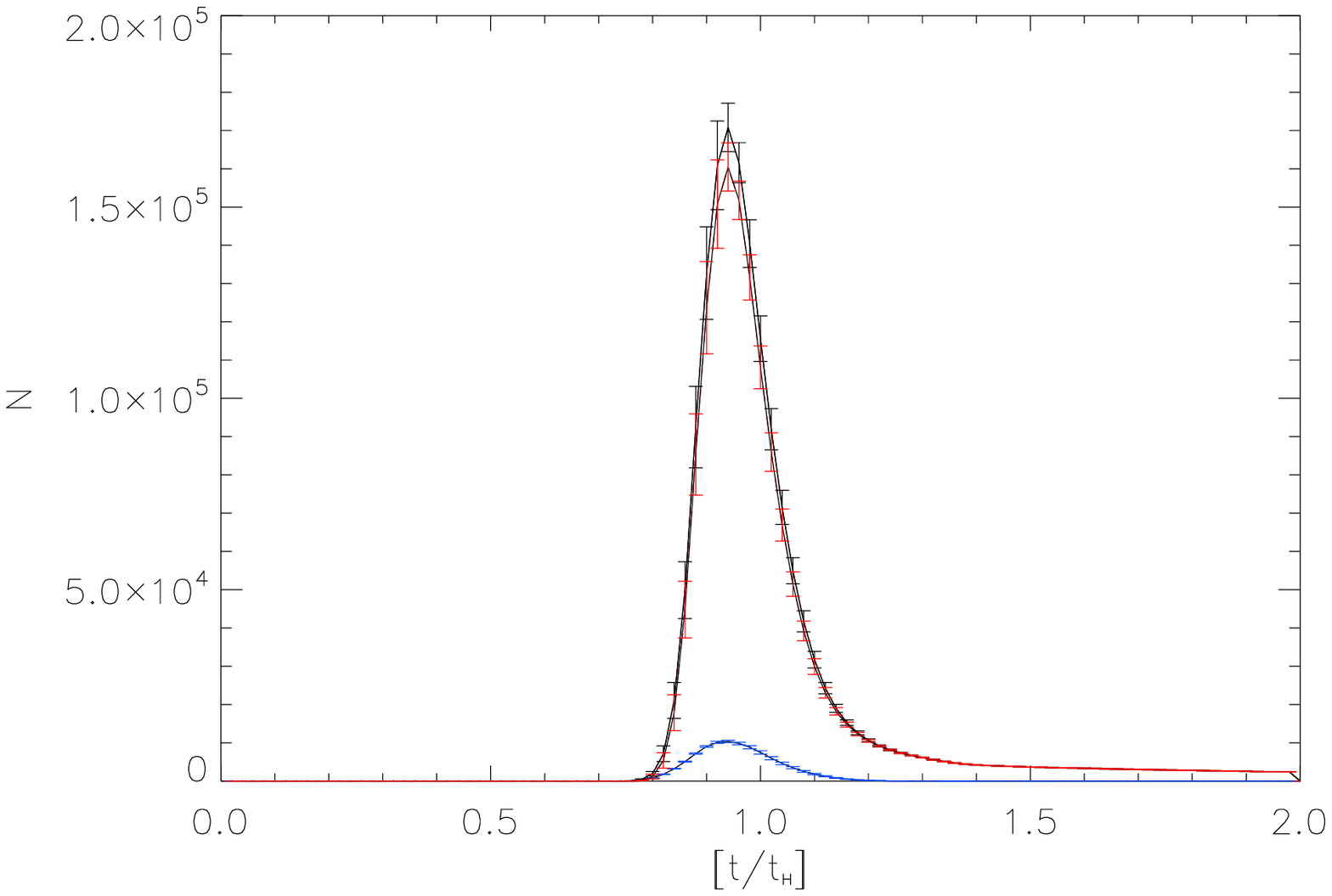} \\
\end{array}$
\caption{The effect of interstellar colonisation speed on strategy
  success.  As the colonisation speed increases from $\chi=10^{-5}$
  (top), $\chi=10^{-4}$ (middle) and $\chi=10^{-3}$ (bottom), the
  parasitic strategy becomes more and more favourable, eventually
  dominating. \label{fig:signal_col}} 
\end{center}
\end{figure}

\begin{figure}
\begin{center}
\includegraphics[scale=0.5]{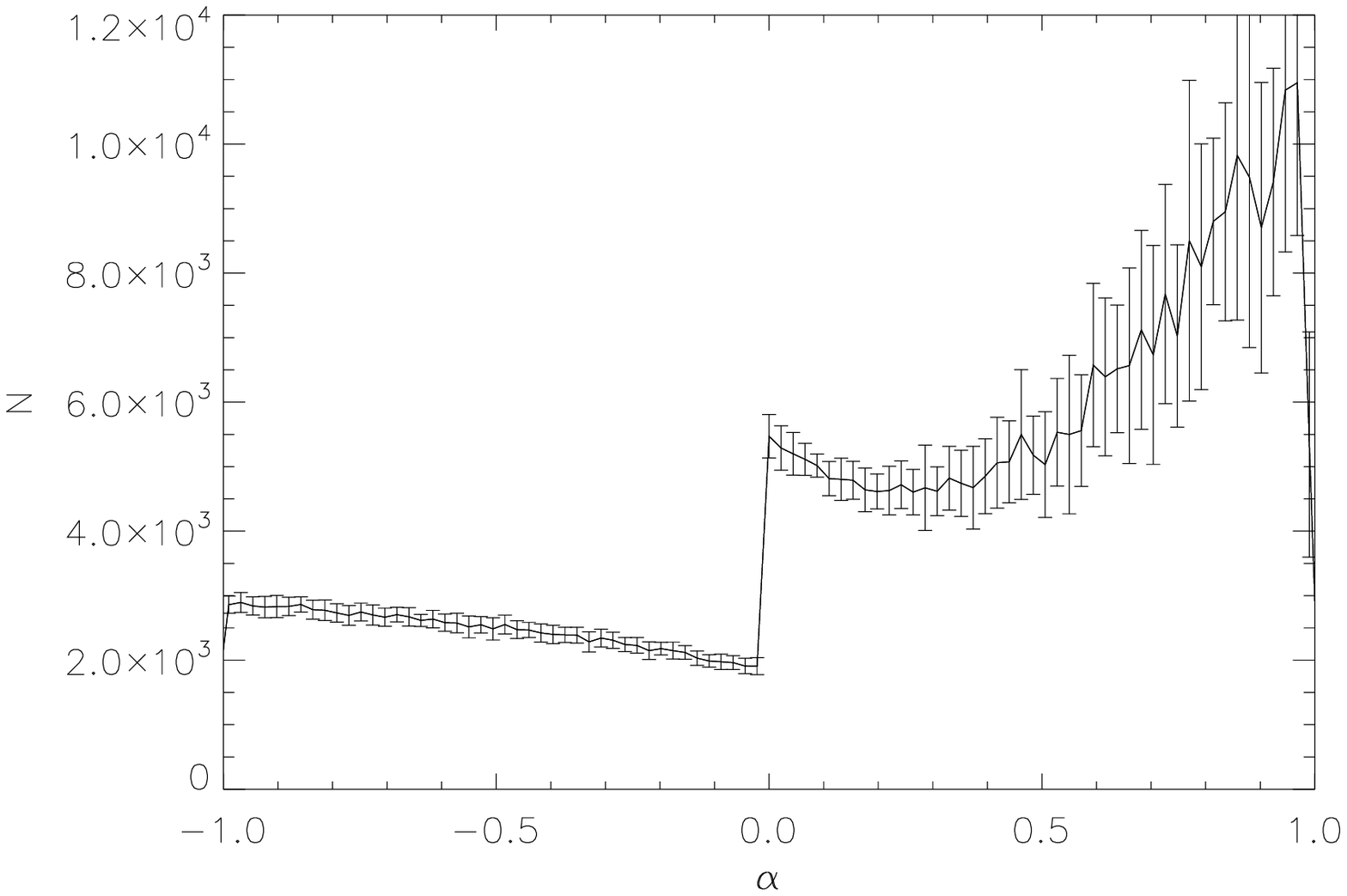} \\
\caption{The effect of interstellar colonisation on civilisation
  strategy (in the case where $\chi=10^{-4}$).  More parasitic
  civilisations are more effective at colonisation, so the
  distribution in $\tilde{\alpha}$ becomes skewed towards more
  positive values. This effect increases with increasing
  $\chi$.  \label{fig:alpha_1e_4}} 
\end{center}
\end{figure}

We therefore model three colonisation scenarios, corresponding to
$\chi=10^{-5},10^{-4},10^{-3}$.  The effect on the number of
intelligent civilisations can be seen in Figure \ref{fig:signal_col}.
Even at low colonisation speed, parasitism is markedly more
successful, increasing its peak numbers by approximately a factor of
two.  The total number of civilisations also receives a boost to its
peak, steepening the transition from low $N$ to high $N$.  The middle
plot of Figure \ref{fig:signal_col} (where $\chi=10^{-4}$) shows
parasitism now the most numerous colonisation strategy in the Galaxy.
The curve becomes even narrower, suggesting that in a moderate-traffic
Galaxy, the likelihood of large numbers of planets colonised by the
same civilisation grows.  In the high velocity case,  parasites
dominate, and mutualism is reduced in peak value by a factor of three.   

We can see this effect in the distribution of $\tilde{\alpha}$ (Figure
\ref{fig:alpha_1e_4}, for the case $\chi=10^{-4}$).  As the
colonisation velocity increases with $\tilde{\alpha}$, more parasitic
civilisations will have a better chance of succeeding at interstellar
colonisation, resulting in a heavy bias towards positive
$\tilde{\alpha}$.  

\section{Discussion \label{sec:discussion}} 

\noindent We note that the assumptions we have made regarding
civilisation behaviour are exactly that: assumptions.  We selected the
virulence paradigm as it is a set of self-consistent assumptions
with a strong biological motivation with which to consider a highly
speculative subject.  We discuss our results presuming
our model is correct, but we acknowledge its simplicity, and that there is no way to
confirm its verisimilitude at this time.

In general the models strongly suggest that a \textbf{K}-strategy
(mutualism) is the preferred strategy for civilisations to engage in
at an interplanetary scale, and that an \textbf{r}-strategy
(parasitism) is only viable provided interstellar travel (at a minimum
speed) is obtainable.  If this minimum colonisation speed is easily
achieved, then parasitism should be the dominant model of
extraterrestrial colonisation and, all things being equal, would lead
to a rapid colonisation of the galaxy.   

Based on the above model, presuming our assumptions are correct, the
results are potentially informative for solving the Fermi Paradox.
That humanity has so far been unable to detect the presence of
extraterrestrial civilisations suggests one of several possibilities: 

\begin{enumerate}
\item Our assumptions about the evolution of extraterrestrial life and intelligence are wrong, or 
\item Sufficiently rapid interstellar travel (such that a parasitical approach would rapidly colonise the galaxy) is difficult to achieve, perhaps impossible.
\item Parasitic civilisations are unable to survive long enough to begin interstellar colonisation.  
\end{enumerate}

\noindent If the models presented in this work are not predicated on
incorrect assumptions, the indication is that the mutualistic approach
is the norm.  The energy budget for colonisation ships to reach the
critical velocity for successful parasitism is simply too high;
civilisations in general will focus on the slow (but successful)
colonisation of its immediate solar system and only rarely engage in
interstellar colonisation and ``empire-building'' \citep{empire}. 

There is, of course, the need for further refinement of what is still
a basic model.  Assigning all civilisations a constant
$\tilde{\alpha}$ is almost certainly an oversimplification. The
possible interactions between parasitical civilisations and
mutualistic civilisations, or the interaction between rival
parasitical civilisations are interesting questions that should also
be addressed; and insight from immunology may be useful for modelling
these in the future.  The experience of invasion biology, in
terrestrial systems, offers mixed insight here as well.  In general,
\textbf{K}-strategy species which have specialised for their
environment (in a largely mutualist fashion) are resistant to invasion
provided that the environment as a whole remains unchanged -
\citet{Williamson1996} discusses various factors affecting the success
of invasions (see also \citealt{Allen2001} for an intriguing
discussion on ecological resilience and invasion).  Invasive species
are unable to outcompete the established species, except in the event
of disturbance (either natural, such as hurricances, or artificial,
such as human-induced ecological degradation).  To what degree this
applies to interstellar colonisation is an open question.
Additionally, we have not fully addressed the influence that
terraforming capacities may have on colonisation rates (with our
current model being limited to the colonisation of already habitable
worlds, i.e. of naturally occurring hosts, as opposed to the conscious
creation of hosts by the symbiont).   

Also, we have restricted our discussion to one particular behavioural
paradigm inspired by biological dynamics - many others could be viable
as well.  A good example is the spread of disease or fungal parasites
amongst plant populations \citep{Otten2004, Gibson2006}.  Approaches
based on percolation theory show that a critical threshold exists for
widespread infection, and that this threshold depends on the
properties of the system, including its spatial geometry.  This is
quite a similar finding to our own in this case - our analogous
threshold is defined by the colonisation velocity. 

A further modelling problem relates to questions of ethics.  While it
is not possible to speculate on the cultural values that may develop
in extraterrestrial civilisations, it is arguable that in order to
develop sufficiently to the degree where interplanetary colonisation
is possible, societies might require a high degree of cooperation
\citep{DeSousa2010}.  The implications of the benefits of mutualism, within
solely the interplanetary region, may mean that civilisations almost
invariably adopt a mutualist phase before they are capable of engaging
in interstellar colonisation.  It is not clear whether, having
developed a mutualist civilisation, the civilisation will then enter
into path dependency (and stay mutualist) or if the civilisation may
change to a more parasitical phase.  Would a mutualist civilisation
even have an interest in pursuing interstellar colonisation, with the
exception of escaping its indigenous stellar collapse?  If mutualist
civilisations only rarely engage in interstellar colonisation, and
parasitical civilisations are unviable, would this explain the Fermi
Paradox?   

These findings - and the questions they raise - have direct relevance
to the challenges currently facing humanity.  \emph{Ceteris paribus},
the findings of our model (cautiously) suggest that it is in
humanity's interest to adopt a more mutualistic relationship with our
host, the Earth (cf \citealt{Berenbaum1999, Speth2009}).  For example,
investing strongly in conservation biology and ecological restoration
are important components of any mutualist approach, the fruits of
which will be important for developing controlled terraforming
techniques for our immediately reachable planets.  Additionally,
investment in reversing human-induced climate change (and the
associated social and economic changes required, as well as
technological improvements) will be required; in short, the focus of
humanity should be on realising sustainable development; indeed, we
believe the ``civilisation as symbiont'' model provides the groundwork
on which to establish a scientifically measurable definition of
‘sustainable development’.  While we are beginning to identify
extrasolar planets in the habitable zone of their parent star
\citep{Batalha2013}, the technology required to reach them is beyond
our current capacity.  Within our own solar system there are no other
fully habitable planets, although the potential for partially
habitable zones in various locales exist, and there are a variety of
niches from which humans could extract useful resources.  For example,
water is thought to exist in liquid form on Europa
\citep{ocean_europa} and Enceladus
\citep{plume_enceladus,plume_enceladus_2}, as well as ices in
environments such as the lunar regolith \citep{Anand2011}.  While not
offering habitats for humanity, they may provide assistance and supply
to space-borne habitats in orbit. 

Finally, two important caveats must be noted:

\begin{enumerate}
\item Our calculation of the critical colonisation velocity for
  parasite success is a sensitive function of the total number of
  civilisations in the Galaxy.  While the stellar and planetary
  parameters of the simulations are reasonably well-constrained, the
  poorly constrained biological parameters will have a deleterious
  effect on the accuracy of the critical velocity. 
\item It should be noted that absence of evidence is not evidence of
  absence - the lack of observed ETIs developing unsustainably is not
  ironclad proof that this strategy is always unsuccessful. 
\end{enumerate}

\section{Conclusions \label{sec:conclusions}}

We have conducted Monte Carlo Realisation (MCR) simulations, adopting
a ``civilisation as symbiont'' model to determine the fate of
civilisations attempting interplanetary and interstellar colonisation.
Civilisations grow and evolve from non-intelligent organisms in a
synthetic, statistically representative Milky Way, containing stars
and planets with properties constrained by observations and theory.
Each civilisation is randomly assigned a set of parameters which
dictates their subsequent colonisation strategy.  Our aim was to study
which colonisation strategy is preferred: whether civilisations are
more successful adopting a mutualist, \textbf{K}-strategy, or a
parasitic, \textbf{r}-strategy \citep{MacArthur1967}.

Our results indicate that in the absence of efficient interstellar
colonisation, mutualist strategies are more successful than parasitic
strategies, much in the same way that \textbf{K}-strategy species
succeed over \textbf{r}-strategy species in isolated environments such
as remote islands.  However, if interstellar colonisation can proceed
at a sufficiently rapid rate, we find that parasitic species will
eventually be favoured.  The critical colonisation velocity for
parasites to dominate is close to the current velocity records
established by humanity both with manned and unmanned spacecraft.
This colonisation velocity has an extremely high energy budget per ton
of mass, suggesting that it is difficult to achieve.  This would
suggest therefore that mutualism is the dominant colonisation strategy
in the Galaxy, and therefore interstellar visitors to the Solar System
are uncommon.

While this does not preclude radio signals travelling between
inhabited worlds, we believe that our model offers a partial solution
to the Fermi Paradox in respect of face-to-face contact.  Our results
are consistent with the findings of invasion biology (and to a lesser
extent the colonial experiences of humanity).  The simplicity of the
model begs further refinement, and we believe later versions of the
model will be able to fold in results from ecology, biogeography,
symbiosis and immunology.  While we acknowledge that the model
  presented in this paper is speculative, simplistic and dependent on a number
  of assumptions, we believe that its general approach, and insights,
  provides a means of framing speculation on broad trends in
  colonisation behaviour.  It is our hope that this research may provide the
  foundation for additional research in applying such
  insight from theoretical ecological modelling.

\section{Acknowledgements}

\noindent This work has made use of the resources provided by the
Edinburgh Compute and Data Facility (ECDF,
http://www.ecdf.ed.ac.uk/). The ECDF is partially supported by the
eDIKT initiative (http://www.edikt.org.uk).  We thank Dr. Neil
McRoberts of UC Davis for comments on an early version of the paper. 

\bibliographystyle{mn2e} % (must include a bibliography style)
\bibliography{ijastarling}

\end{document}